\documentclass[a4paper,12 pt]{article}

\usepackage[utf8x]{inputenc}
\usepackage {graphicx}
\usepackage{amsmath}
\usepackage{amssymb}
\usepackage{authblk}
\usepackage{mdframed}
\usepackage[switch]{lineno} 

\usepackage{setspace}

\definecolor{light-gray}{gray}{0.70}
\definecolor{dark-gray}{gray}{0.30}

\begin{document}

\title{On the interplay of speciation and dispersal: \\An evolutionary food web model in space}
\author{Korinna T. Allhoff \thanks{Corresponding author: allhoff@fkp.tu-darmstadt.de}} 
\author{Eva Marie Weiel \thanks{schoenwald@fkp.tu-darmstadt.de}} 
\author{Tobias Rogge \thanks{rogge@fkp.tu-darmstadt.de}}
\author{Barbara Drossel \thanks{drossel@fkp.tu-darmstadt.de}} 

\affil{Institute of Condensed Matter Physics 
\\Darmstadt University of Technology \\Hochschulstra\ss e 6, 64289 Darmstadt, Germany}

\maketitle

\begin{abstract}
We introduce an evolutionary metacommunity of multitrophic food webs on several habitats coupled by migration. 
In contrast to previous studies that focus either on evolutionary or on spatial aspects, we include both and investigate the interplay between them. 
Locally, the species emerge, interact and go extinct according to the rules of the well-known evolutionary food web model proposed by Loeuille and Loreau in 2005. 
Additionally, species are able to migrate between the habitats. 
With random migration, we are able to reproduce common trends in diversity-dispersal relationships: 
Regional diversity decreases with increasing migration rates, whereas local diversity can increase in case of a low level of dispersal. 
Moreover, we find that the total biomasses in the different patches become similar even when species composition remains different. 
With adaptive migration, we observe species compositions that differ considerably between patches and contain species that are descendant from ancestors on both patches. 
This result indicates that the combination of spatial aspects and evolutionary processes affects the structure of food webs in different ways than each of them alone.
\end{abstract}

 

\onehalfspacing
\newpage
\section{Introduction}
Classical food web models represent an idealization of real ecosystems that focuses on feeding relationships as the most important type of interaction and that considers populations as well mixed and homogeneous in space. 
Typically, such models include nonlinear differential equations that capture the growth and loss terms of population dynamics, and a simple stochastic algorithm for generating network structures with realistic features, such as the niche model \cite{Nische2000} or the cascade model \cite{Cohen1985, Cohen1990}. 
They provide a static, mean field description, integrating the feeding relationships across the whole spatial extent of the system and ignoring temporal changes in the composition of the network due to species turnover.

In order to go beyond mean-field models, various approaches have been taken to include spatial structure or species turnover in food web models. 
If space has the structure of discrete habitats, one obtains  "networks of networks". 
The outer network represents the spatial landscapes consisting of several habitats, the connections between them representing possible routes for dispersal. 
A chain topology of habitats results for instance for a river with barrages, and a ring of habitats can occur along island shores. 
More complex spatial networks might represent archipelagos,  or a system of waterbodies connected by streams and canals. 
The inner networks describe localized food webs on these habitats, the connections between species representing feeding relationships. 
The need to study such spatially extended food webs has been highlighted recently by several authors \cite{Hagen2012, Gonzalez2011, Amarasekare2008, Leibold2004}. 
Most studies of spatial ecosystems concentrate on simple topologies of the inner network, 
such as food chains \cite{Calcagno2011} or small food web motifs of two \cite{Jansen2001,deRoos1998}, three \cite{Reichenbach2007, Koelle2005, Holt2002, Blasius1999} or four \cite{Ristl2014} species in space. 
So far, there exist few investigations of larger food webs in space, both empirical \cite{Legrand2012, Logue2011, Presley2010, Cottenie2003}, and theoretical \cite{Haegeman2014, McCann2005, Mouquet2002, Mouquet2003, Wilson1992}.
Moreover, all of the mentioned studies focus on spatial aspects under the assumption that the species composition is static. 

On the other hand, studies addressing species turnover typically neglect spatial aspects. 
During the last years, several models were introduced that include evolutionary dynamics (for references see next paragraph). 
On a time scale much slower than population dynamics, new species, which are modifications of existing species, are added to the system. 
They can be interpreted either as invaders from another, not explicitly considered spatial region, or as arising from a speciation process. 
Population dynamics then determines which species are viable. In contrast to static models such as the niche model, the food web structure is not put in by hand, but emerges from the interplay between population dynamics and species addition. 
Evolutionary food web models can therefore give insights into the conditions under which complex network structures can emerge and persist in face of ongoing species turnover. 
They are thus fundamentally different from species assembly models, which have been studied for a longer time and which are based on a fixed species pool from which species are added to a smaller habitat. 

In 2005, Loeuille and Loreau \cite{LL2005} introduced the probably simplest successful evolutionary food web model. 
In contrast to other well-known evolutionary food web models, like for example the matching model \cite{Rossberg2006, Rossberg2008} or the webworld model \cite{Caldarelli1998, Drossel2001, Drossel2004}, 
which describe a species by a vector of many abstract traits, a species in this model is specified only by its body mass. 
The feeding relationships are determined by differences in body mass. 
A version with gradual evolution was studied by  Br\"annstr\"om et al in 2011 \cite{Braennstroem2011}. 
Ingram et al \cite{Ingram2009} extended the model to include an evolving feeding range,  and Allhoff and Drossel \cite{Allhoff2013} also considered a version with an evolving feeding center. 
These extensions make the model very similar to the evolving niche model \cite{Guill2008}, 
where the niche value can be equated with the logarithm of the body mass, and where also these three parameters are evolved. 
In contrast to the simpler model by  Loeuille and Loreau \cite{LL2005}, these models need additional ingredients 
that prevent evolution from running to extremes, such as adaptive foraging or restrictions on the possible trait values \cite{Allhoff2013}.

Recently, several authors emphasized that combining the spatial and the evolutionary perspective on ecosystems 
is essential for better understanding coexistence and diversity \cite{Logue2011, Urban2006, Urban2006b, Urban2008}. 
It is well known that including a spatial dimension in evolutionary models enables the coexistence of species or strategies that would otherwise exclude each other \cite{Szabo2007}. 
This is due to the formation of dynamical waves in which the competitors cyclically replace each other, or to the formation of local clusters that cannot easily be invaded from outside. 
However, these studies are usually limited to two or three species. 
A recent study of a larger system was published in 2008 by Loeuille and Leibold \cite{Loeuille2008}, 
who investigated a metacommunity food web model with two plant and two consumer species on a patchy environment, where one of the plant species has evolving defense strategies. 
The authors demonstrated the emergence of morphs that could only exist in a metacommunity due 
to the presence of dispersal highlighting the fact that the combination of space and evolutionary processes yields important new insights. 

In this paper, we study the combined effect of space and evolution on food webs consisting of many species on up to four trophic levels. 
We use the model of Loeuille and Loreau \cite{LL2005}, placing it on several habitats that might represent lakes, islands or a fragmented landscape, and that are coupled by migration. 
The results are "evolutionary networks of networks". 
By varying migration rules (undirected, directed, diffusive, adaptive, dependent on body mass), the time of migration onset (at the beginning or after local food webs have evolved), and the number and properties of habitats (2 or 8 habitats, equivalent or differing with respect to simulation parameters), we investigate many different scenarios. 

With diffusive migration, our results agree qualitatively with diversity-dispersal relationships from empirical studies \cite{Sax2003} and from other theoretical metacommunity studies \cite{Mouquet2002, Mouquet2003, Urban2006b}. 
Low migration rates lead to an increased diversity in the local habitats, and high migration rates lead to homogenization of habitats and hence to a decreased regional diversity. 
For a chain of eight habitats coupled by diffusive migration, we find that migration leads to equal biomasses in the habitats, even when the species composition of neighboring patches is very different. 
With adaptive migration we obtain networks that differ strongly in their species composition but that do not show increased local diversity.


\newpage
\section{Model and Methods}
\subsection{The Model by Loeuille and Loreau on one habitat}
The model by Loeuille and Loreau \cite{LL2005} includes population dynamics on the one hand 
and the introduction of new species via modification of existing species on the other. 
Because such "mutation" events are very rare, population dynamics typically reaches an attractor before the introduction of a new species.
Thus, ecological and evolutionary time scales can be viewed as separate. 

\bigskip
Population dynamics is based on the body mass $x_i$ of a species $i\in \{1,...,n\}$ as its only key trait.
Species are sorted such that body mass increases with index number.
Production efficiency $f$ and mortality rate $m$  scale with body masses 
according to the allometric relations $f(x_i) = f_0x_i^{-0.25}$ and $m(x_i) = m_0x_i^{-0.25}$.
The population dynamics of species $i$ with biomass $N_i$ is given by
\begin{align}
 \frac{dN_i}{dt} &= f(x_i) \sum_{j=0\;}^{i-1} \gamma_{ij} N_iN_j \	&\text{(predation input)}   \notag \\
&- m(x_i) N_i & \text{(mortality)}   \notag \\
&- \sum_{\;j=1}^{n} \alpha_{ij} N_iN_j  & \text{(competition)}  \notag \\
&- \sum_{j=i+1}^{n} \gamma_{ji} N_iN_j & \text{(predation loss)} \label{eq:popdyn} 
\end{align}
with 
\begin{equation}
 \gamma_{ij} = \gamma(x_i-x_j) = \frac{\gamma_0}{s\sqrt{2\pi}}\exp\left(\frac{-(x_i-x_j-d)^2}{s^2}\right)
\end{equation}
describing  the rate with which predator $i$ consumes prey $j$, and with 
\begin{equation}
\alpha_{ij}=\left\{ 
\begin{array}{l} \alpha_0 
\quad \mbox{ if }  |x_i-x_j|\leq \beta\, \\
0 \quad \;\;\mbox{ else\, }
 \end{array}
\right.  
\end{equation}
describing the competition strength. 
The parameters $\gamma_0$, $d$, $s$, and $\beta$ are the integrated feeding rate, the preferred body mass difference between predator and prey, the width of the feeding niche, and the competition range. 

Energy input into the system is provided by an external resource of ``body mass'' $x_0=0$ and total biomass $N_0$, which is subject to the dynamical equation
\begin{align}
 \frac{dN_0}{dt} &= I-eN_0 - \sum_{i=1}^{n} \gamma_{i0} N_i N_0  \notag \\
 &+ v \left(\sum_{j=1}^{n}m(x_i)N_i  
 +  \sum_{i=1}^{n} \sum_{j=1}^{n} \alpha_{ij} N_iN_j 
 + \sum_{i=1}^{n} \sum_{j=0}^{i-1} \left(1-f(x_i) \right) \gamma_{ij} N_iN_j \right) \; {.}  \label{eq:recycling}
\end{align}
The terms represent constant input of inorganic nutrient, nutrient outflow, consumption by the basal species, and recycling of a proportion $v$ of the biomass loss due to mortality, competition and predation.

Starting from a single ancestor species of body mass $x_1=d$, the food web is gradually built by including evolutionary dynamics in addition to the population dynamics. 
New species are introduced with a "mutation" rate of $10^{-6}$ per unit mass and unit time. 
The new, ``mutant'' species has a body mass that deviates by at most 20 percent from the body mass $x_i$ of the ``parent'' species 
and is drawn randomly from the interval $[0.8x_i, 1.2 x_i]$. 
Body masses can not get arbitrary big or small due to the constant value of the preferred predator-prey body mass difference, 
which leads to the typical network structure outlined at the beginning of the results section. 
Loeuille and Loreau set the initial biomass of the mutant to $10^{-20}$, which is also the extinction threshold for all species.
Therefore only mutants with a positive growth rate can add successfully to the system or even replace other species, those with negative growth rates go extinct immediately.

\bigskip
We used the C language to perform our computer simulations and chose all parameters as in \cite{LL2005} except for the initial biomass and extinction threshold, which was $10^{-6}$ instead of $10^{-20}$. 
Simulations were usually run for $4\cdot 10^8$ or $2\cdot 10^7$ time units and with the original mutation rate of $10^{-6}$ or an increased mutation rate of $2\cdot 10^{-5}$, respectively. 
These changes do not influence the results, but allow for faster calculations. 
For comparison, the generation time of the ancestor species with body mass $x_i=d=2$ and mortality rate $m_0=0.1$ 
is of the order of $\frac{1}{m(x_i)}=\frac{x_i^{0.25}}{m_0} \approx 12$ time units.

Loeuille and Loreau demonstrated that their model is able to produce a large variety of robust network structures dependent on the competition and feeding parameters \cite{LL2005}. 
If not indicated otherwise, we use a fixed set of parameters 
($d=2$, $\frac{s²}{d}=0.5$, $f_0=0.3$, $m_0=0.1$, $\gamma_0=1$, $\alpha_0=0.1$, $\beta=0.25$, $I=10$, $e=0.1$, $v=0.5$)
in order to concentrate on the effects generated by the spatial landscape. 
With this parameter set, networks of approximately $16-18$ species emerge.

\subsection{Rules used for two habitats coupled by migration}
We chose a simple diffusion approach to describe migration between two equivalent habitats. 
Each species can have two populations, one on each habitat. 
To describe feeding and competition interactions that take place within one habitat, we use the evolutionary model as explained above. 
For each population of species $i$ on each habitat $h$ we add a migration term to equation (\ref{eq:popdyn}): 
\begin{equation}
 ... - \underbrace{\mu_{i,h\rightarrow h'}N_{i,h}}_{\text{migration from}\;h\;\text{to}\;h'} + \underbrace{\mu_{i,h'\rightarrow h}N_{i,h'}}_{\text{migration from}\;h'\;\text{to}\;h} 
\qquad
\left[h, h' \in [1,2], \; h\neq h'\right]  \label{eq:migration}
 \end{equation}
Here, $\mu_{i,h\rightarrow h'}$ and $\mu_{i,h'\rightarrow h}$ are the migration rates of species $i$. 
We do not include loss terms, which means that all  biomass leaving one habitat appears on the other. 
The resource is supposed to be confined to its habitat and does not migrate, $\mu_{0,h\rightarrow h'}  = \mu_{0,h'\rightarrow h} = 0$. 

We consider diffusive migration, 
where the migration rates are treated as constants $\mu_{i,h\rightarrow h'} =\mu$ for all $h,h',i$ with $ \mu \in \left(10^{-1}, 10^{-2}, ..., \leq 10^{-6}\right)$. 
We also investigate the variant with directed migration, for which $\mu_{i,h\rightarrow h'} = 0$ if $h'>h$.  
For high values of $\mu$, dispersal occurs on a similar time scale as population dynamics; for instance the mortality rate of the ancestor species with $x_1=2$ is $m_0x_1^{-0.25}\simeq 0.085$.
Moreover, we investigate the case of allometrically scaled migration rates, 
 $\mu_{i,h\rightarrow h'} = \mu_{i,h'\rightarrow h} = \mu \cdot x_i$, where species with larger body masses migrate faster.

All these cases are analyzed in two versions. 
In the first version, the migration rates are zero during the initial build-up of the networks. 
Migration sets in only after the networks have fully emerged. 
This mimicks situations in which separate ecosystems become coupled, for example by the building of canals between waterbodies, or by the formation of land bridges. 
In the second version, migration sets in at the beginning of the simulation so that both habitats coevolve.
We also discuss the case where migration is switched off after some time.

In systems with migration, the extinction threshold must be treated differently than in the original model. 
Due to emigration, the biomass of new mutants can initially fall below the extinction threshold, even when the new species is viable. 
We therefore first applied population dynamics after each ``mutation'', and only after the new population equilibrium was reached did we remove species that were below the extinction threshold. 
Interestingly, population dynamics always went to a fixed point; we never saw periodic oscillations or chaotic attractors. 

We repeated all simulations several times in order to make sure that the results are generic. 
In the most intensely studied case of diffusive migration between two equivalent habitats, we performed more than 450 simulation runs. 

\subsection{Variants}
In order to get a deeper understanding of the system, 
we also performed more than 160 simulations of three scenarios where either the spatial landscape or the migration is designed in a more complex way. 

\begin{itemize}

\item Inhomogeneous systems: \\
In real ecosystems, habitats can differ with respect to temperature, nutrients, resources, size, etc. 
To implement such inhomogeneous systems, we performed simulations 
where the habitats differ either in their competition strength ($\alpha_{0,h=1}= 0.1$ and $\alpha_{0,h=2}= 0.02$) or in their competition range ($\beta_{h=1}= 0.25$ and $\beta_{h=2}= 0.125$).
Both competition parameters have an important influence on the resulting network structures \cite{LL2005, Allhoff2013}. 
Smaller values of the competition strength $\alpha_0$ imply less intraspecific competition and therefore bigger, but fewer populations 
because the energy provided by the resource can only support a certain total biomass.  
Smaller values of the competition range $\beta$ imply less competitive exclusion, allowing for more species per trophic level.
The migration in this variant is implemented as undirected diffusion as explained above. 

\item Chain of habitats: \\
In another variant with a more complex spatial landscape we discuss a chain of 8 equivalent habitats. 
Again, we choose undirected diffusive migration with a constant migration rate $\mu_{i, h\rightarrow h\pm1} = \frac{\mu}{2}$ between neighboring habitats $h \in [1,8]$. 
Hence, the additional migration term for species $i$ on habitat $h$ is 
\begin{equation}
   ...+\; \underbrace{\frac{\mu}{2} (N_{i,h+1}+N_{i,h-1})}_{\text{immigration}} \;- \underbrace{\mu N_{i,h}}_{\text{emigration}} \label{eq:chain}\;.
\end{equation}
We choose closed boundary conditions, i. e. we set $N_{i,0} = N_{i,1}$ and $N_{i,9}=N_{i,8}$ in equation \ref{eq:chain}. 

The increased number of habitats leads to a significantly increased program runtime. 
To keep it within a reasonable limit, we decreased the value of the feeding range to $s=0.5$ (only for this variant). 
This leads to better adapted, but fewer predators and hence to a decreased system size of approximately $12-15$ species per isolated habitat \cite{Allhoff2013}.

\item Adaptive migration: \\
Up to now, migration is diffusive and hence based on random movement. 
However, especially for higher developed species that can evaluate their current situation and possibly follow their prey or avoid competitors, this might be too simple. 
In this variant, we go back to 2 equivalent habitats, but instead of diffusive migration, 
we analyze two versions of adaptive migration, where the migration rates of the species are dependent on their current growth rates $G_{i,h}$. 
\begin{itemize}
 \item Type I:  A species $i$ emigrates from  habitat $h$ if its population size in that habitat  is currently decreasing, 
 \begin{align*}
  \mu_{i,h\rightarrow h'} = \left\{\begin{array}{cl} c \cdot G_{i,h} & \mbox{if } G_{i,h}<0\\ 0 & \mbox{else }\end{array}\right. \quad .
 \end{align*}
 \item Type II: Migration is directed into the habitat with better local conditions, 
  \begin{align*}
  \mu_{i,h\rightarrow h'} = \left\{\begin{array}{cl} c \cdot \left(G_{i,h'}-G_{i,h} \right) & \mbox{if } \left(G_{i,h'}-G_{i,h} \right)>0\\ 0 & \mbox{else }\end{array}\right. \quad .  
 \end{align*}
\end{itemize}
We varied the value of the proportionality factor $c$ over three orders of magnitude from $10^{-2}$ to $10$.  

\end{itemize}


\newpage
\section{Results}
\subsection{Diffusive migration between two equivalent habitats}


\begin{figure}[ht]
  \centering
 \includegraphics[]{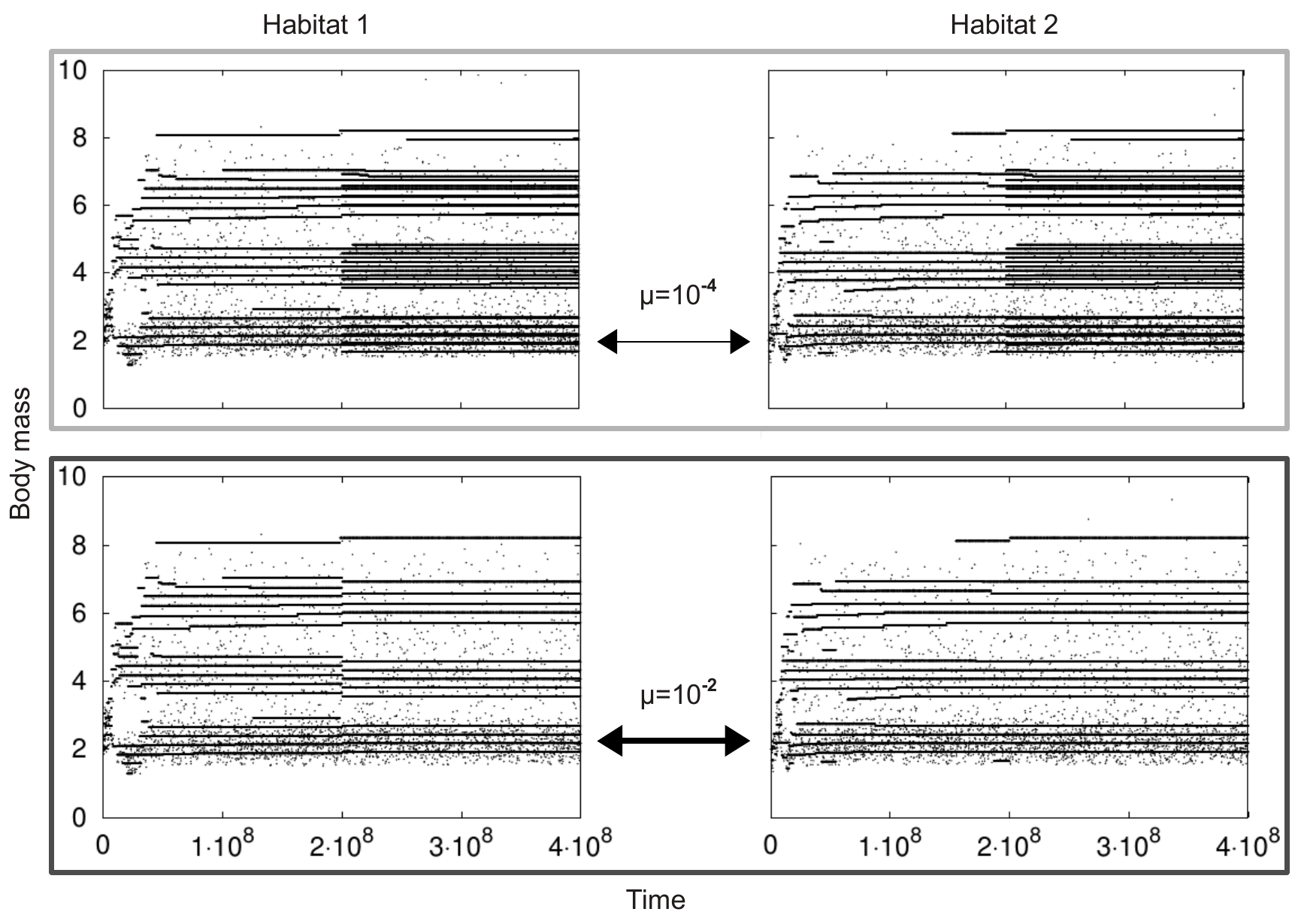}
 \caption{Evolving body masses in two habitats, simulated with the model of Loeuille and Loreau \cite{LL2005}. 
 Migration between the habitats with $\mu=10^{-4}$ (top line) or $\mu=10^{-2}$ (bottom line) starts at $t=2\cdot10^8$ after the initial build-up of the networks. 
 The resulting networks are shown in the top line of fig. \ref{fig:summary}.}
 \label{fig:Ung_BM}
\end{figure}

\begin{figure}[p]
  \centering
  \includegraphics[]{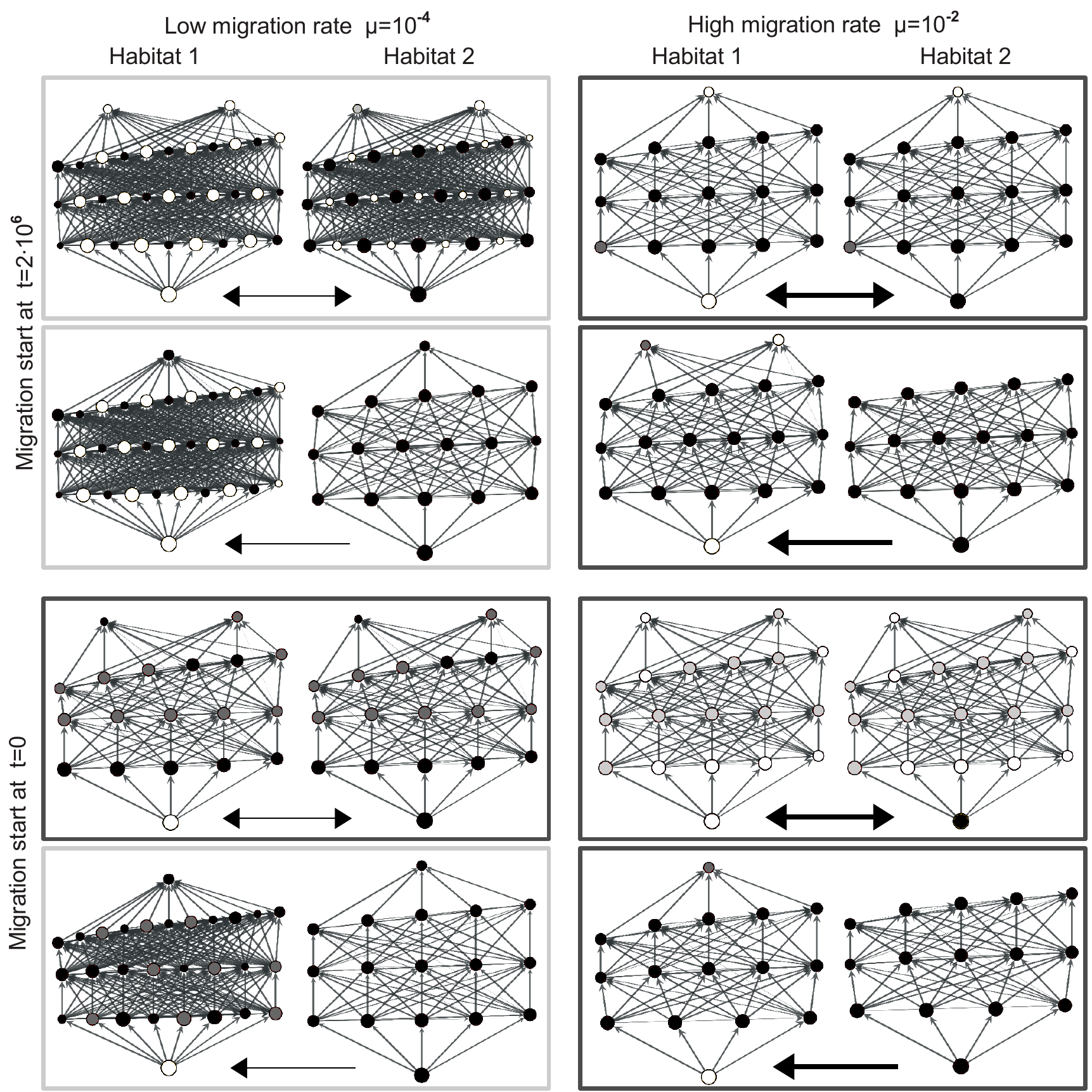}
 \caption{Example networks on two habitats coupled by weak (left panels) or strong (right panels) migration. 
 The arrows between the networks indicate the direction of migration. 
 The vertices represent species, with the radius scaling logarithmically with biomass density. 
 The colors of the species represent their habitat of origination: 
 White (black) species are natives of habitat 1 (2) and originated there, too. 
 Light (dark) gray species originated in habitat 2 (1), but are descendants of a white (black) species. 
 The arrows between species represent feeding links, with the width scaling logarithmically with the attack rate. 
 The vertical position of a species represents its trophic position, which is the average, weighted trophic position of its prey, plus one ("flow-based TL" \cite{Williams2004a}). 
 Time evolution of the networks in line 1 is shown in fig. \ref{fig:Ung_BM}.
 The color of the frames around the networks indicates the two possible outcomes. 
 Light gray frames: networks with small additional populations (outcome 1). Dark gray frames: similar / identical networks (outcome 2). 
 Network visualizations given in this paper are based on \emph{graph-tool} (http://graph-tool.skewed.de).
 \label{fig:summary}}
\end{figure}

Starting from a single ancestor species, the evolutionary model of Loeuille and Loreau \cite{LL2005} goes first through a  period of strong diversification, 
and then the network structure stabilizes and assumes a regular pattern. 
Fig. \ref{fig:Ung_BM} shows the body masses of all species occurring during two exemplary simulations in two habitats that are initially isolated. 
Species with a body mass of $x_i \approx l\cdot d$ consume species with a body mass of approximately $(l-1)\cdot d$ and form the $l^{th}$ trophic level. 
These trophic levels are blurred for large values of the niche width $\frac{s^2}{d}$ (not shown) and clearly separated for small values. 
To avoid competition, species keep a minimum body mass difference of $\beta$, allowing for more species on a trophic level when $\beta$ is smaller \cite{Allhoff2013}. 
If two species are so similar in body mass that they compete with each other (e.g. parent and mutant species), only one of them survives. 
Although the realizations in the two habitats are based on the same set of parameters, they show slightly different structures due to different sets of random numbers. 

At time $t=2\cdot 10^8$, undirected diffusive migration between the two habitats according to equation (\ref{eq:migration}) sets in. 
Dependent on the migration rate, we identified two major outcomes, here and in the following marked by a light or dark gray frame:  
\begin{enumerate}
 \item 
 \begin{mdframed}[linecolor=light-gray, linewidth=2pt]
 For small migration rates (e.g. $\mu=10^{-4}$, see top line in fig. \ref{fig:Ung_BM}), migrants have small additional populations in the foreign habitat, leading to an approximately doubled number of species per habitat. 
 The resulting network structures are thus combinations of the isolated networks. 
\end{mdframed}\medskip
 \item 
 \begin{mdframed}[linecolor=dark-gray, linewidth=2pt]
 In case of a high migration rate (e.g. $\mu=10^{-2}$, see bottom line in fig. \ref{fig:Ung_BM}), native species become displaced by invaders. 
 The resulting networks are very similar or even identical in the two habitats. 
\end{mdframed}\medskip
\end{enumerate}

In both cases, the outcome is reached soon after the onset of migration. 
Since the immigrants arrive in a habitat where the network is already completely developed, every niche is already occupied and all immigrants have to compete with native species. 
If migration rate is small, the immigrants' gain in biomass due to migration and feeding interactions becomes soon canceled by competition losses, and the immigrant populations stay small. 
As soon as migration is switched off, these small populations vanish again (not shown). 
With a higher migration rate, some immigrants can establish themselves against their competitors and displace native species. 
Again, all species (invaders and natives) tend to keep a body mass difference of $\geq\beta$ to minimize their competition loss. 
Since all species from one habitat and especially from one trophic level coevolved together, they are in this respect well matched to each other. 
As a consequence, often complete levels are replaced. 


In fig. \ref{fig:summary}, the resulting network structures of several simulation runs are shown. 
The first line corresponds to the simulations shown in fig. \ref{fig:Ung_BM}. 
The colors of the species represent their habitat of origin. 
White species are natives to habitat 1 and black species are natives to habitat 2. 
If a black species migrates into habitat 1 and has a mutant there, this mutant is colored dark gray. 
Light gray species have analogously originated in habitat 2 but are descendants of a white species from habitat 1. 

We also analyzed scenarios where migration is only allowed in one direction or where migration starts at the beginning of the simulation. 
Directed migration (fig. \ref{fig:summary}, line 2) leads to similar results for the immigration habitat as undirected migration. 
The network structure in the emigration habitat depends on the migration rate. 
The migration loss can be formally regarded as an increased mortality term for all species in habitat 2. 
In case of very low migration rates, this term is negligible and leaves the network structure of habitat 2 unchanged. 
In case of high migration rates, a significant amount of biomass leaves the emigration habitat per unit time, leading to the extinction of one or more species from the upper levels. 

If migration starts at the beginning of the simulation (fig. \ref{fig:summary}, line 3), both resulting networks are identical and only either black and dark gray or white and light gray species occur. 
The first successful mutant that replaces its ancestor species in its home habitat is also able to migrate to the other habitat and displace also the ancestor's population there. 
Every subsequent mutant is a descendant from this first mutant and finds identical conditions in both habitats, leading to identical networks. 

Line 4 of fig. \ref{fig:summary} shows results of directed migration during the whole simulation time. 
We observe a combination of the explained effects. 
All species are descendants of the first successful invader from habitat 2 and therefore either black or dark gray. 
For a low migration rate, we observe again small additional populations in habitat 1 (outcome 1).
For a high migration rate, the networks are mostly identical (outcome 2) except for the top level.

In order to gain a general overview of the influence of the migration rate, we
varied the value of $\mu$ over several orders of magnitude and performed 48 simulation runs with $\mu \in( 10^{-1}, 10^{-2}, ... ,10^{-6})$. 
Larger or smaller migration rates influence the time needed until the system reaches a new fixed point after a mutation event, 
but the resulting network structures do not provide any new insights besides the two explained outcomes. 
Intermediate migration rates lead to a superposition of the two outcomes, where one trophic level contains additional small populations corresponding to outcome 1
and another trophic level is replaced by species from the other habitat corresponding to outcome 2. 
We observed essentially the same effect for body-mass dependent migration rates, 
with the migration rate being proportional to the body mass (another 120 realizations, data not shown). 
Since now species on higher levels had larger migration rates, they were  more often replaced, while lower levels showed more often additional populations. 
However, since in this model all body masses are of the same order of magnitude, body-mass effects are only minor.


\begin{figure}[ht]
 \includegraphics[]{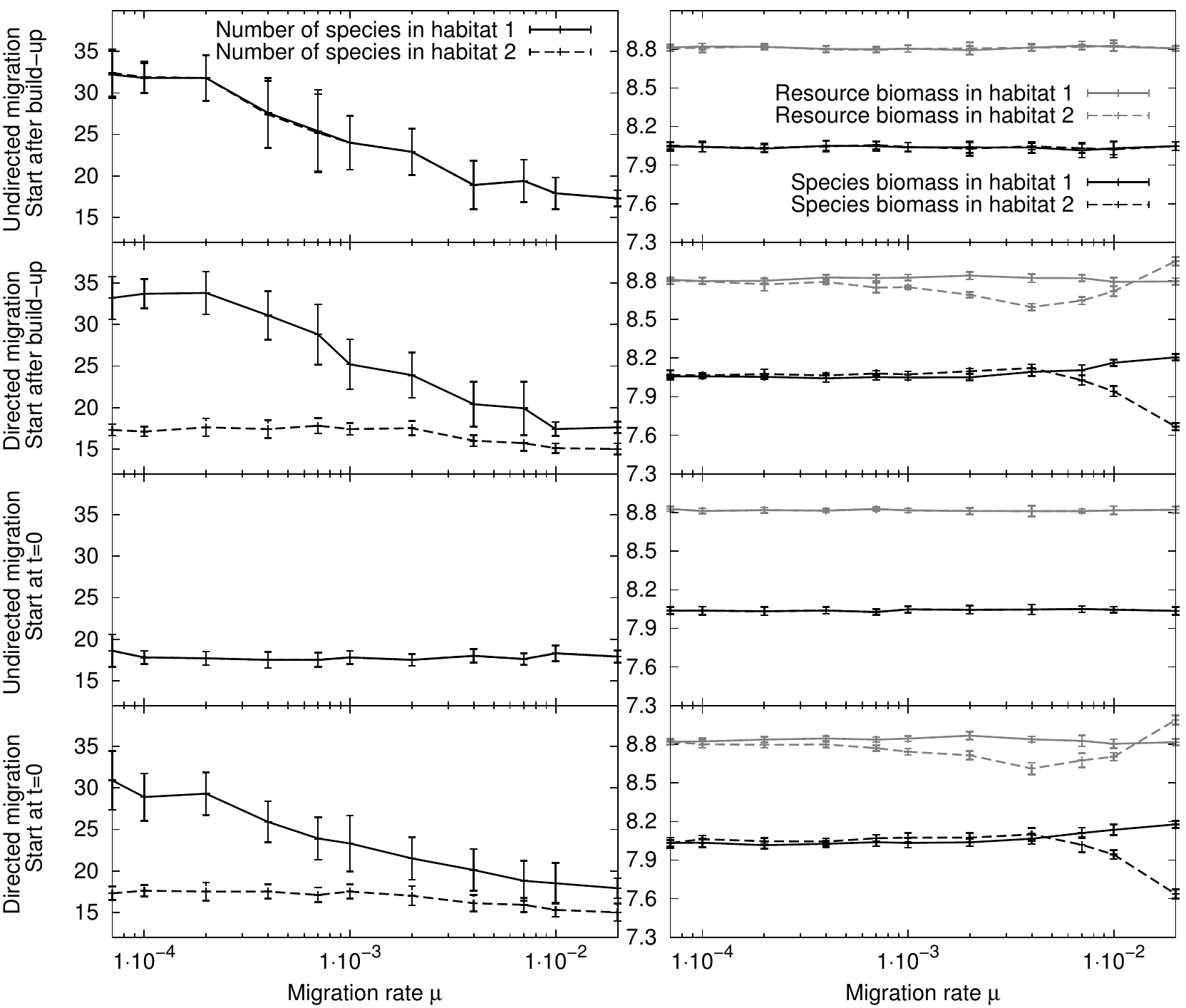}
 \caption{Network size and total biomasses of resources and all other species in dependence of the migration rate $\mu$. 
 Each data point represents the average and standard deviation of 10 simulation runs with different random numbers. 
 Isolated habitats show the same results as realizations with undirected migration starting at the beginning of the simulation (line 3).}
 \label{fig:SB}
\end{figure}

\begin{figure}[ht]
 \centering
 \includegraphics[]{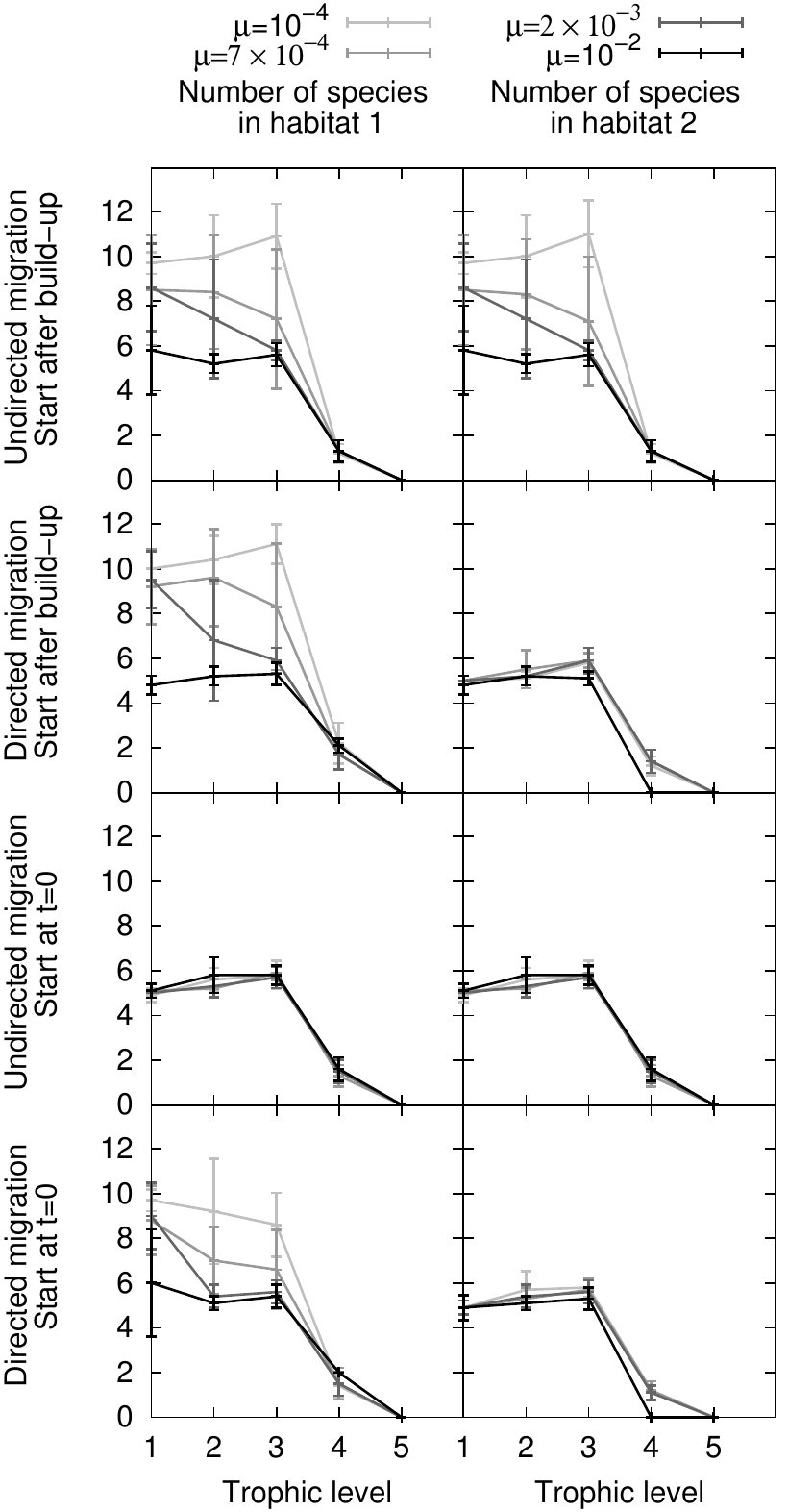}
 \caption{Number of species per trophic level (rounded to nearest integer values) for simulations of $4$ different scenarios and for $4$ values of the migration rate $\mu$. 
 Each data point represents average and standard deviation of 10 realizations with different random numbers. 
 Isolated habitats show the same distribution as realizations with undirected migration starting at the beginning of the simulation (line 3).}
 \label{fig:STL}
\end{figure}

\begin{figure}[ht]
  \centering
 \includegraphics[]{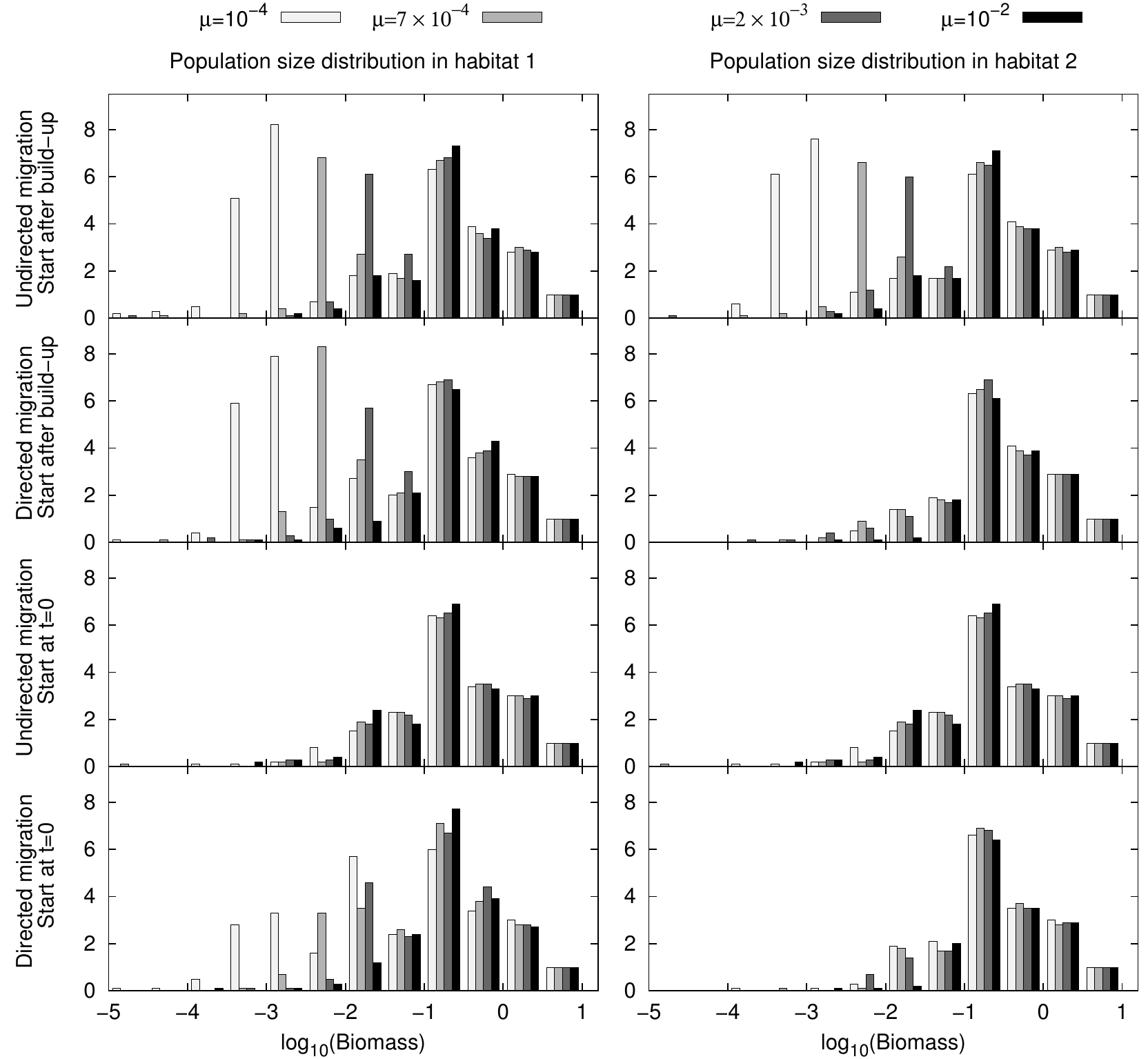}
 \caption{Distributions of populations sizes for simulations of $4$ different scenarios and for $4$ values of the migration rate $\mu$. 
 Each column represents an average over 10 realizations with different random numbers. 
 In all realizations, the resource had by far the biggest population ($N_0\approx 8.8$). 
 Isolated habitats show the same distribution as realizations with undirected migration starting at the beginning of the simulation (line 3). }
 \label{fig:Biomass}
\end{figure}

To understand the transition from small to larger migration rates in more detail, we performed more than 400 simulation runs with $2 \cdot 10^{-2}\geq \mu \geq 7 \cdot 10^{-5}$. 
In fig. \ref{fig:SB} we show that the transition between the described two outcomes is smooth and covers approximately two decades of migration strength. 
With undirected migration starting after the initial build-up (line 1), each species has populations in both habitats, so that the number of species per habitat is identical. 
In case of a high migration rate, not only the species number, but the whole networks are identical (outcome 2), 
whereas in case of a rather small migration rate, the network size is approximately doubled due to small additional populations (outcome 1). 
However, even if the species composition strongly depends on the migration rate, the total biomasses of the resources and the total biomass of the species do not (see top line of fig. \ref{fig:SB}). 
It is even nearly identical for both habitats and shows very small variations across the realizations that differ only in the set of random numbers. 

The situation is different with directed migration (fig. \ref{fig:SB}, line 2). 
The number of species in habitat 1 shows a similar smooth transition from many additional populations to the displacement of native species, 
whereas habitat 2 accommodates a rather constant number of species. 
Only in case of high migration rates, species from upper trophic levels become extinct due to the migration losses, as explained above. 
This leads to a non-monotonous dependence of the biomasses on the migration rate, according to the following top-down trophic cascade \cite{Oksanen1981}: 
For $\mu\approx 3\cdot 10^{-3}$, all species in the fourth trophic level of habitat 2 are extinct due to migration losses, 
so that species in the third level experience no predation pressure. 
Hence, even despite their own migration losses, they can have big populations and exert a high pressure on the second level. 
Due to the subsequent reduction of population sizes in the second level, species in the first level also experience a reduced predation pressure, have big populations 
and exert a high pressure on the external resource, which is observed as a reduced resource biomass. 
For even higher migration rates, also species from the third level in habitat 1 become extinct due to migration losses, the total biomass of the species decreases and the resource recovers. 

With a migration start at the beginning of the simulation and undirected migration (fig. \ref{fig:SB}, line 3), we observe identical networks, as explained above. 
The results do not depend on the migration rate, and are identical with  results from simulations of isolated habitats (not shown). 
If directed migration is active during the build-up of the networks (fig. \ref{fig:SB}, line 4), 
we observe in principle the same effects as when migration sets in after the initial build-up. 
However, some immigrants occasionally find an empty niche as long as the build-up is not yet completed. 
They do not have to compete with natives and can establish themselves, reducing the number of additional small populations. 
Note that this figure does not show the fact that all species are black or dark gray (see line 4 of fig. \ref{fig:summary}).

The number of species per habitat can also be interpreted as local diversity. 
In case of undirected migration, local and regional diversity are identical, since every species has populations on both habitats. 
In case of directed migration, the local diversities differ, and the local diversity of habitat 1 is again the regional diversity. 
Hence, we observe that low migration rates can lead to an increased local diversity, whereas high migration rates lead to a decreased regional diversity. 

The distribution of the species per trophic level reveals more details about the transition from small additional populations to the displacement of native species, as shown in fig. \ref{fig:STL}. 
Here, light gray represents small migration rates and dark gray represents high migration rates. 
Higher trophic levels show the transition at smaller values of the migration rate $\mu$ than lower trophic levels: 
In case of $\mu=2\cdot10^{-3}$, nearly no additional populations were observed on the third level, but many on the second and even more on the first level. 
We ascribe this to the fact that in this model species on higher trophic levels have smaller populations than species on lower trophic levels, 
and therefore exert a lower competition pressure on the invaders. 
However, as also observed in fig. \ref{fig:SB}, the error bars are biggest for intermediate migration rates, 
indicating that dependent on the random numbers single simulations might deviate from this trend. 
This is due to the above mentioned fact that whole trophic levels (not only single species) show either outcome 1 or outcome 2, 
which leads to an increased number of possible network structures.

The population sizes of the invaders of outcome 1 depend on the migration strength, see fig. \ref{fig:Biomass}. 
In those scenarios that show the discussed transition between the outcomes and for low migration rates (light gray columns, $\mu=10^{-4}$), 
we observe a bimodal frequency distribution of population sizes. 
In addition to the population sizes that also occur in outcome 2 for higher migrations rates (black columns, $\mu=10^{-2}$), 
also peaks at smaller population sizes occur, which correspond to the additional populations of outcome 1.
For higher migration rates, these peaks shrink and shift to larger populations sizes, in agreement with the smooth transition shown in the previous figures.

\clearpage
\subsection{Variants}

\subsubsection{Inhomogeneous system}
\begin{figure}[ht]
\includegraphics[]{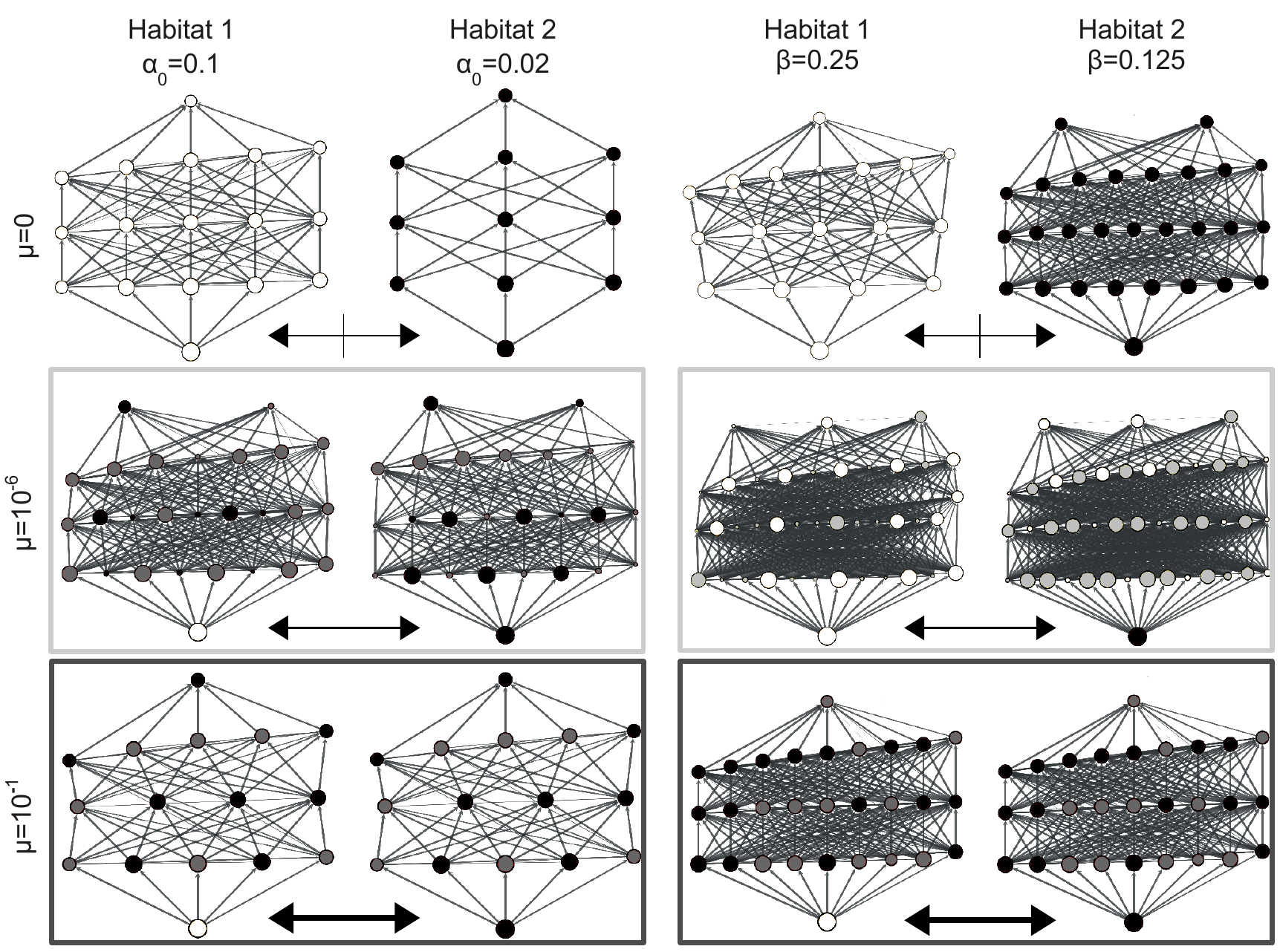}
\caption{Undirected diffusive migration between two habitats that differ in the competition strength $\alpha_0$ (left) or in the competition range $\beta$ (right). 
Migration is absent (top) or starts at $t=0$ with $\mu=10^{-6}$ (middle) or $\mu=10^{-1}$ (bottom). 
For more explanations see caption of fig. \ref{fig:summary}.
\label{fig:comp}}
\end{figure}

Fig. \ref{fig:comp} shows example networks resulting from simulations with different competition parameters.
Due to the different environments on the two habitats, different network structures emerge in the two patches if they are uncoupled (see upper line in fig. \ref{fig:comp}). 
A decreased competition strength $\alpha_0$ leads to bigger, but fewer populations (left) 
and an increased competition range $\beta$ leads to more competitive exclusion and therefore to smaller networks (right) \cite{Allhoff2013}. 

When coupled by weak migration, the resulting networks look like a superposition of the isolated networks, see middle line in fig. \ref{fig:comp}. 
Each species exists in both habitats resulting in an increased number of populations. 
However, the population of one species is large in one habitat and small in the other, like the additional populations in outcome 1. 
Counting only the big populations, one recognizes the network structures of the isolated habitats. 
A stronger migration link (bottom line) leads to identical networks consistent with outcome 2. 
Similar network structures but with mixed colors can be obtained when migration starts after the networks have developed (not shown).

\subsubsection{Chain of habitats}
\begin{figure}[ht]
\centering
  \includegraphics[]{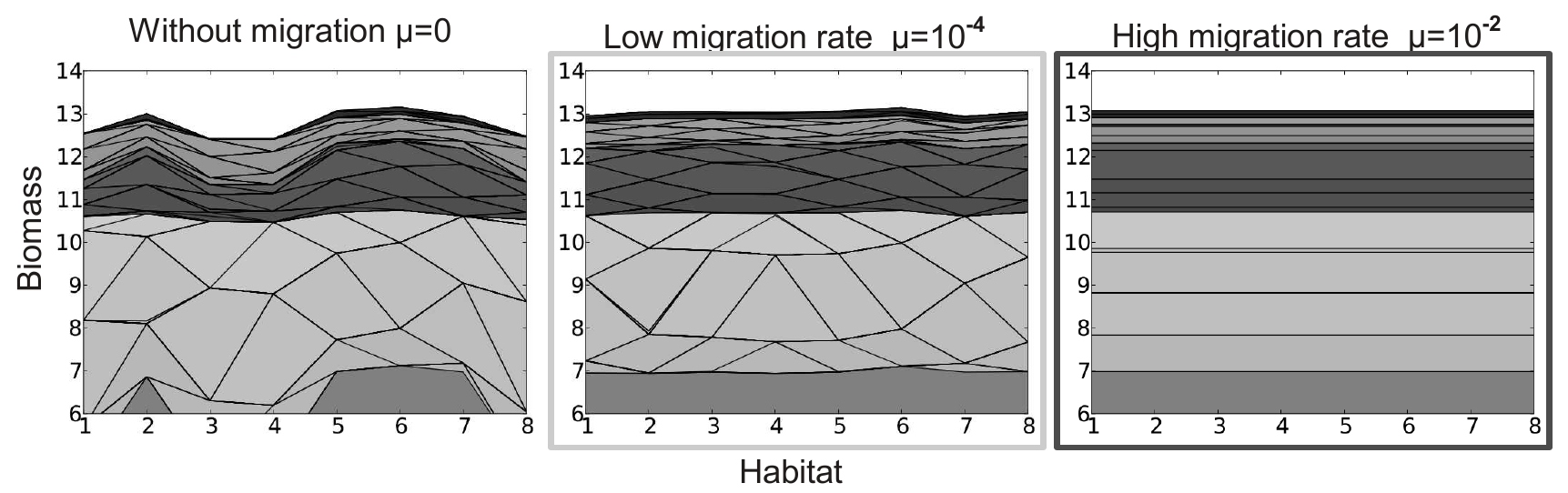} 
 \caption{Simulation outcome for undirected diffusive migration along a chain of 8 habitats. Migration starts after the initial build-up of the networks. 
  Given is the biomass distribution of the resulting food webs before migration sets in (left panel) 
  and after the emergence of new structures due to weak migration (middle panel) or strong migration (right panel). 
  Networks of habitat 4 and 5 with weak migration are shown in fig. \ref{fig:chainnetworks}. 
 \label{fig:chain}}
\end{figure}

\begin{figure}[ht]
\centering
\includegraphics[]{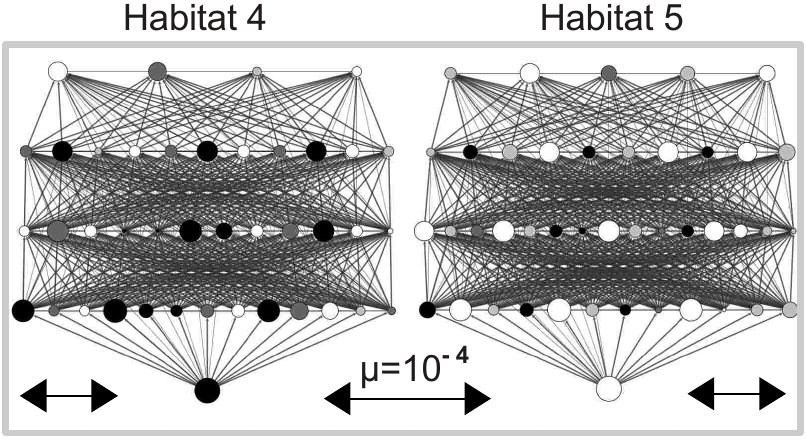}
\caption{Example networks on a chain of habitats weakly coupled by migration. 
The color code is different from the networks of fig. \ref{fig:summary}. 
Black (white) vertices: Species that originated in habitat 4 (5).
Dark (light) gray vertices: Species that originated further on the left (right).}
\label{fig:chainnetworks}
\end{figure}

As an example of an extended spatial landscapes, we discuss a chain of habitats. 
In the left panel of fig. \ref{fig:chain}, the biomass distributions of eight isolated habitats are shown. 
Due to the randomly chosen mutant body masses, each network consists of a unique species composition. 
Some compositions seem to be more favorable than others in the sense that the total amount of biomass (of species and resources) is larger. 

After all eight networks have fully emerged, migration is switched on. 
In the case of weak migration (middle panel) the situation is again similar to outcome 1, where the immigrating species can not establish themselves. 
Their populations stay much smaller than the natives and survive only due to the continuous migration into the habitat.
These additional populations are too small to be visible in fig. \ref{fig:chain}, but are obvious in the example networks of habitat 4 and 5 shown in fig. \ref{fig:chainnetworks}. 
The color code in this figure is different to the previous figures. 
Black species with big populations in their native habitat 4 have small populations in their neighboring habitat 5 and vice versa. 
Also shown are small populations from the habitats further on the left in dark gray or further on the right in light gray. 
These additional populations have a major effect on the system concerning the biomass distributions. 
Even if the invaders can not establish themselves, they provide a continuous energy flow between the habitats. 
As a consequence, all biomass distributions equalize. 
This result does not depend on the recycling loop in equation (\ref{eq:recycling}), but occurs also when the recycling loop is switched off. 

Stronger migration (right panel of fig. \ref{fig:chain}) leads again to outcome 2. 
Immigrating species can establish themselves and displace natives. 
Transiently very large networks occur while all species migrate in both directions and are present in many habitats at once. 
Then, by and by, the most favorable species composition (i.e., the one with the largest total biomass) displaces others and the resulting networks are identical. 
However, this process takes much longer as with only two habitats. 

We also discussed the same scenario of 8 habitats with a migration start at time $t=0$ (not shown). 
Then, all networks coevolve. 
If the mutation rate is still so small that a successfully mutant can spread over all habitats before a new mutant emerges, 
identical networks emerge, in consistency with the corresponding scenario of undirected diffusive migration between two habitats (line 3 in fig. \ref{fig:summary}). 

Just as for the case of 2 patches, we made sure that these results are generic by performing several simulation runs and by using more values of the migration rates. 

\subsubsection{Adaptive migration}
\begin{figure}[ht]
 \centering
 \includegraphics[]{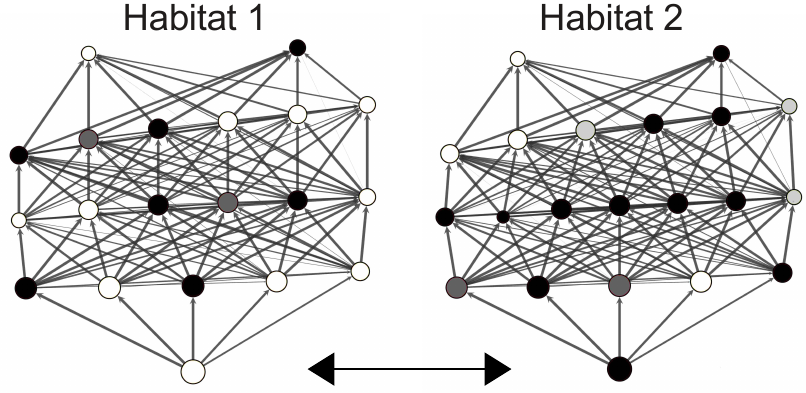} 
 \caption{Resulting networks of a simulation with undirected adaptive migration starting after the initial build-up at $t=2\cdot 10^{6}$. 
 The migration rate of a species is proportional to the difference of its growth rates in the two habitats (Type 2), here with a factor $c=10$. 
 For more explanations see caption of fig. \ref{fig:summary}.
 \label{fig:adaptive}}
\end{figure}

Finally, we discuss two types of adaptive behavior. 
With a migration start after the initial build-up of the networks, both types show in principle the same behavior. 
Since the system is most of the time near a fixed point with zero growth rate for all species, migration can only occur rarely, when the system is disturbed by a successful mutant. 
If the mutant replaces another species, this species has a negative growth rate and hence can migrate to the other habitat, where it can possibly replace an already existent species and establish itself.
This makes the two habitats different from each other and can lead to completely different species compositions as shown in the example networks in fig. \ref{fig:adaptive}.
However, the general structure of the networks remains unchanged and is the same as that of isolated networks. 
The probability that a replaced species can successfully invade the other habitat increases with the proportionality factor $c$ of the migration rate, in consistency with the previously discussed cases. 
With $c=1$ only few species invade the other habitat successfully and with even smaller values of $c=0.1$ or $c=0.01$ the networks are basically isolated with very rare successful invasions.

The replacement of one species by a similar mutant or invader is only a  small perturbation of the whole system. 
Every prey or predator of the dying species experiences small fluctuations in its population size during the replacement. Consequently, other species can also migrate into the other habitat. 
However, additional mini-populations like in outcome 1 can not occur, since these disturbances and hence the migration rates are very small and occur transiently. 
Thus, these species cannot establish themselves on the other habitat and do not occur in figure \ref{fig:adaptive}. 

With a migration start at the beginning of the simulation, two mostly identical networks emerge (not shown). 
They differ at most by one single species, which just emerged in one habitat with a monotonously increasing population size and therefore did not have the opportunity to migrate.
This is consistent with the corresponding scenario of diffusive migration between two habitats (line 3 in fig. \ref{fig:summary}).


\section{Discussion}
We have studied an evolutionary food web model on several habitats.
Locally, species emerge, interact and go extinct according to the evolutionary food web model of Loeuille and Loreau \cite{LL2005}. 
Additionally, they migrate between habitats according to diffusive or adaptive dispersal. 
The computer simulations start with one species in each habitat. 
Migration may occur from the beginning, so that the food webs in the different habitats coevolve, or it may occur later after the local food webs have become established. 
Usually, our computer simulations show one of two frequent outcomes: 
Either the local food webs of the different habitats become mostly identical and have a similar structure and size as in an isolated system (outcome 2), 
or the local food webs differ with respect to the main species in each trophic layer but include small populations from the neighboring habitats 
that are sustained by ongoing immigration but cannot displace the native occupant of a niche (outcome 1). 
Which of the two outcomes occurs depends mainly on the migration rate and the time of migration onset. For intermediate migration rates, the two outcomes can combine, with part of the trophic levels being identical in the two patches, and with other levels being different and showing small populations from the other patch.
Furthermore, we find that even for outcome 1 the total biomasses in the habitats become very similar to each other. 
To our knowledge, this result has not yet been observed in other studies.
When the habitats are not equivalent because migration is directed or because the model parameters are different, the food webs can differ with respect to species number and population sizes. 
When migration occurs only under certain circumstances (for instance when the growth rate of a population is negative), the species composition of neighboring habitats can become very different.

Generally, the food webs in the habitats show the regular structure that is characteristic of the model by Loeuille and Loreau \cite{LL2005}. 
They are clearly structured into distinct levels, and each level consists of niches separated by a body mass difference that is equal to the competition range. 
After the initial stage of strong diversification, species turnover becomes very slow, new mutants are only rarely successful and very similar to their parent species, and the food web structure is very stable. 
The model can therefore not reproduce the sometimes catastrophic effect of species invasions into natural ecosystems, 
where alien species may find such good local conditions that their populations grow explosively 
leading to the extinction of prey or competitors species and a cascade of secondary extinctions \cite{Courchamp2003, Fritts1998}. 
In the model by Loeuille and Loreau, species differ only with respect to body mass. 
If an invader successfully replaces a similar species, it has automatically the same predators and the same prey and hence the same function in the food web. 
Thus, the displacement of a species leaves the overall network structure unchanged. 
Once the initial build-up of the network is complete, all viable niches are occupied and stay occupied so that secondary extinctions can not occur. 

The complexity of real ecosystems of cause exceeds by far the complexity of this model \cite{Driscoll2009}. 
The interactions between species do not only depend on body mass but also on many other species traits and also on environmental factors. 
The latter show considerable variations in space and time, causing species to change continuously, 
as for example implemented in another model presented by Loeuille and Leibold in 2008 \cite{Loeuille2008b}. 
After all, not only a changing environment, but also the local feedback between species and their environment makes the food webs and the migration behavior highly diverse \cite{Loeuille2014}.

The simple model studied in this paper highlights those effects of migration on evolving ecosystems that occur already when only few traits are taken into account and when habitats are equivalent. 
The two main outcomes described above are widely observed in empirical and theoretical studies. 
Sax et al. \cite{Sax2002} investigated invasions and extincts of land birds and vascular plants on oceanic islands. 
They found that for land birds the number of naturalizations of nonnative species is roughly equal to the number of extinctions, 
whereas for vascular plants species richness has increased by about a factor of two. 
The authors give several possible explanations for this behavior, one of them posits that nonnative species have become established because they are competitively superior to natives. 
If applied to birds, this would correspond to outcome 2. 
The increase in the number of plant species is similar to outcome 1. 
In our model, such an increase could be explained by ongoing immigration sustaining additional populations that would go extinct otherwise. 
However, this appears to be an unlikely explanation for oceanic islands, where the increased diversity is rather being attributed to an increased variety in local habitats, including those created by man \cite{Sax2002}.

Other theoretical studies of metacommunities also show the two types of outcomes. 
Mouquet and Loreau \cite{Mouquet2002, Mouquet2003} studied the effect of migration on local and regional diversity in a non-evolving metacommunity. 
Their model was later extended by Urban to contain adaptive phynotypic variation in the reproductive rates of 20 competing species inhabiting 20 heterogeneous patches \cite{Urban2006b}. 
Without dispersal, all communities are unique and isolated, leading to a low local diversity and a high regional diversity. 
With a low or intermediate level of dispersal, regional diversity remains unchanged whereas local diversity increases due to immigration from neighboring communities. 
This corresponds to outcome 1, where the total number of species does not change after the onset of migration, but where the local number of species is approximately doubled (in a 2-patch system). 
Mouquet and Loreau predicted that higher levels of dispersal lead to homogenization of the metacommunity and hence to decreasing local and regional diversities, in consistency with our outcome 2. 

Also many other studies suggest that local and regional diversity react differently to changes in the spatial landscape and dispersal \cite{Sax2003}. 
Haegemann and Loreau \cite{Haegeman2014} extended the investigations by analyzing different dispersal rates for resources and consumers 
leading also to local consumers with regional resources or regional consumers with local resources. 
The latter corresponds to our case of body-mass dependent migration rates, 
where species with small body masses in the lower trophic levels experience too small migration rates to successfully invade the other habitat, 
whereas the species compositions in the upper trophic levels are homogeneous. 
However, it should be mentioned that in the model by Loeuille and Loreau body mass differences are generally small and metabolic scaling of the migration rates has only weak effects. 
 
Evolutionary species turnover is not necessary for all our results. 
In situations where migration is switched on only after the local food webs have become established, 
the two types of outcomes are also observed when the process of introducing new mutant species is stopped. 
However, in such a case no ``gray'' species would occur (see Fig. 2), which are descendants of immigrants from the other habitat. 
Our computer simulations with adaptive migration, however, yield an outcome that could not be obtained in absence of evolution. 
Since migration rates are dependent on population growth rates, migration occurs only temporally, when the system is disturbed by the emergence of a new mutant. 
This leads to different species compositions in the two habitats. 
Just as the results of Loeuille and Leibold \cite{Loeuille2008} mentioned in the Introduction, 
these findings show that the interplay between space and evolutionary processes gives rise to new phenomena. 
However, we have to admit that the typical outcomes observed in our model do not appear to be very realistic. 
They can probably be attributed to the unusual stability of the model by Loeuille and Loreau. 
In general, adaptive behavior is known to have a considerable stabilizing effect on food web dynamics \cite{Kondoh2003,Guill2008}, 
but it cannot become visible when dynamics is already very stable in the absence of adaptive behavior. 
We expect that a noticeable stabilizing effect will become visible when adaptive migration is combined with more complex and less stable evolutionary food web models. 
Certainly, the study presented in this paper is only a modest beginning of the investigation of evolutionary food web models in space.

\appendix
\footnotesize

\section*{Acknowledgments}
This work was supported by the DFG under contract number Dr300/12-1. \\
We thank Nicolas Loeuille and Daniel Ritterskamp for very useful discussions. 


\end{document}